\documentclass[conference]{IEEEtran}
\IEEEoverridecommandlockouts
\usepackage[T1]{fontenc}
\usepackage{amsmath,amssymb,amsfonts}
\usepackage{algorithmic}
\usepackage{graphicx}
\usepackage{textcomp}
\usepackage{xcolor}
\usepackage{paralist}
\usepackage[backend=biber, style=ieee]{biblatex}
\def\BibTeX{{\rm B\kern-.05em{\sc i\kern-.025em b}\kern-.08em
    T\kern-.1667em\lower.7ex\hbox{E}\kern-.125emX}}

\usepackage[compact]{titlesec}
\titlespacing{\section}{0pt}{0ex}{0ex}
\titlespacing{\subsection}{0pt}{0ex}{0ex}
\titlespacing{\subsubsection}{0pt}{0ex}{0ex} 
\begin{document}

\title{\vspace{-4ex}Security evaluation of 6G-oriented and O-RAN networks through targeted attacks\\
\thanks{This work is partially funded by the European Union. Views and opinions expressed are, however, those of the author(s) only and do not necessarily reflect those of the European Union or Smart Networks and Services Joint Undertaking. Neither the European Union nor the granting authority can be held responsible for them. The XTRUST-6G project (GA no. 101192749) is supported by the 6GSNS and its members.}
}

\author{\\[-5ex]\IEEEauthorblockN{%
Apostolos Valiakos\IEEEauthorrefmark{1}, Dimitrios Psaltis\IEEEauthorrefmark{1}, Alexios Lekidis\IEEEauthorrefmark{1}, Nikolaos Kekatos\IEEEauthorrefmark{1}, Elpiniki Papageorgiou\IEEEauthorrefmark{1}
\vspace*{4pt}}
\IEEEauthorblockA{%
Department of Energy Systems, University of Thessaly, Larissa 41500, Greece\\Email:
\IEEEauthorrefmark{1}\{avaliakos, dpsaltis, alekidis, nkekatos, elpinikipapageorgiou\}@uth.gr}\vspace{-6ex}}

\maketitle

\begin{abstract}
The sixth generation (6G) of networks is supported by increased bandwidth, automation through the use of O-RAN technologies. Simultaneously though autonomous operation leads to an extensive threat surface and sophisticated attacks. Understanding the attack impact, the Tactics, Techniques and Procedures that are used by attackers is extremely challenging, but however is crucial for developing defense and mitigation strategies against them. This paper evaluates a targeted attack chain against a virtualized, 6G-oriented O-RAN testbed in which the F1 midhaul interface is accessible to an attacker and transport-layer protection is not enabled. The attack disrupts the legitimate O-DU association, introduces a rogue O-DU using observed configuration parameters, and intercepts user-plane traffic. The attack was evaluated in a fully virtualized testbed based on Open5GS, the srsRAN Project, FlexRIC, and an Electric Vehicle (EV) charging infrastructure based on OCPP 1.6.
\end{abstract}


\section{Introduction}
The evolution towards 6G networks is expected to rely heavily on openness, virtualization, automation, and intelligence at the edge. Open Radio Access Network (O-RAN) architectures are central to this transition, since they decompose traditional base-station functionality into interoperable components such as the O-CU, O-DU, and O-RU, while exposing standardized interfaces for control, monitoring, and optimization. These properties enable flexible deployment models, lower vendor dependency, and support for emerging vertical applications, including industrial IoT, and EV charging infrastructures.

However, the same openness and disaggregation that make O-RAN attractive also expand the network attack surface. Interfaces that were previously hidden inside proprietary equipment are now transported over IP-based midhaul and fronthaul networks, often across virtualized or software-defined environments. If these interfaces are insufficiently protected, adversaries with access to the transport network may be able to observe signaling traffic, extract configuration parameters, disrupt legitimate associations, or impersonate trusted RAN components. This risk becomes particularly important in private 5G and 6G-oriented deployments, where operators may prioritize low latency, reduced computational overhead, or simplified configuration over strict security enforcement.

Critical infrastructure scenarios amplify the impact of such weaknesses. In EV charging environments, cellular connectivity may support the exchange of Open Charge Point Protocol (OCPP) messages between charging stations and the Charging Station Management System (CSMS). A compromise of the underlying RAN can therefore expose sensitive operational data, authentication identifiers, metering information, and control commands. Such exposure may lead not only to loss of confidentiality, but also to financial fraud, service disruption, and manipulation of energy-consumption records.

This paper investigates the security implications of a targeted attack against a 6G-oriented O-RAN testbed. In particular, we analyze how an adversary positioned inside the IP midhaul network can disrupt the legitimate F1 association between the O-CU and O-DU, extract relevant F1AP parameters, and introduce a rogue O-DU capable of hijacking control and user-plane traffic. The attack is evaluated in an experimental environment based on Open5GS, the srsRAN Project, FlexRIC, and an emulated EV charging environment. The results demonstrate plaintext interception of OCPP payloads at F1-U only in the evaluated configuration, where user-plane ciphering is disabled and the F1 transport is not protected. The concrete contributions of this work are the:

\begin{compactitem} 
    \item Design of a reproducible 6G-oriented O-RAN emulation testbed that separates the O-CU, O-DU, core network, Near-RT RIC, user equipment, and adversarial node across virtualized components.
   \item Analysis of a targeted F1 midhaul attack methodology that combines reconnaissance, transport-layer disruption, and rogue O-DU impersonation.
    \item Demonstration of a practical interception of EV charging traffic by capturing plaintext OCPP messages under a misconfigured or resource-constrained private 6G-oriented network scenario using Null Ciphering.
\end{compactitem} 




\section{Proposed Attack Methodology}

This paper adopts and adapts the methodology developed by Moheddine and Loscri \cite{moheddine2025identifying} for the exploitation of vulnerabilities in the control protocol, at the SCTP level, to compromise legitimate 5G infrastructure. While this work focuses primarily on the mechanics of the DoS attack on the Control Plane (NGAP), our analysis targets the internal F1 (Midhaul) interface of O-RAN. We proceed to the next stage, investigating the complete interception of User Plane (F1-U) traffic by a Rogue O-DU, exploiting the absence of encryption through the Null Ciphering - (NEA0) algorithm \cite{3gpp_ts33501} at the PDCP layer for resource-constrained network devices.  

The attack is applied to a private 6G-oriented network environment for the EV charging infrastructure. Building on previous research \cite{11130009}, we highlight the vulnerability of the OCPP protocol, demonstrating how a network breach translates into immediate financial and operational risk through the interception of JSON payloads, such as OCPP Messages. Specifically, this paper investigates two collaborative methods for compromising the legitimate infrastructure. Initially, ARP Blackholing, as a Man-in-the-Middle attack, follows the patterns of availability attacks established by Shaik et al. \cite{shaik2016practical} and the control-plane vulnerabilities analyzed by Aiello et al. \cite{aiello2026investigating}. Then, a more advanced step is the  SCTP Abort attack. Both techniques were prerequisites for the subsequent configuration of a Rogue O-RAN DU, a spoofing scenario that has been extensively studied by Saedi et al. and Fue et al.~\cite{saedi2020generation, fue2025understanding}.  

The security breach and the implementation of the rogue antenna are achieved through the functional subversion of the interaction between the legitimate O-CU and the O-DU. Specifically, the attack exploits a Protocol-Aware Denial of Service (DoS) attack at the SCTP layer, aimed at disconnecting the legitimate O-DU, followed by a Midhaul Impersonation attack. Consequently, the attacker gains unauthorized access to the F1-C (Signaling) and F1-U (User Data) streams.  

\setcounter{subsubsection}{0}
The attack methodology focuses on undermining the mutual trust that governs O-RAN nodes, achieving penetration through the following phases:  

\begin{compactenum}
    \item \textit{O-DU Level Takeover}: Using ARP Spoofing and SCTP Abort techniques, the connection between the legitimate O-DU and the O-CU is forcibly terminated. Because the attack takes place on the midhaul network, the O-CU incorrectly assumes that the O-DU unit has failed and terminates the session.
    \item \textit{Rogue O-DU Impersonation}: Simultaneously, the srsRAN Project software on Kali Linux launches a Rogue O-DU. The O-CU, and consequently the Open5GS Core, is unable to detect the impersonation, as the Rogue O-DU presents the valid network identifiers (gNB ID, NCI, MCC/MNC) that were previously intercepted via passive monitoring (sniffing). Once the UE (EV Charger) connects to the malicious DU, traffic is automatically routed through the attacker's infrastructure. 
     \item \textit{F1 Interface Manipulation}: In the O-RAN architecture, communication between CU and DU takes place via the F1 interface. Although User Plane encryption is in theory implemented at the PDCP layer within the O-CU, the widespread use of the Null Ciphering algorithm (NEA0) \cite{3gpp_ts33501} practical setups avoid implementing this protection. In many industrial IoT applications, including EV charging infrastructure, the configuration of NEA0 at the User Plane is not uncommon. This practice is adopted by network operators to minimize computational load and processing delays on low cost, resource-constrained cellular modems. Furthermore, in private 5G scenarios, operators rely solely on the network's assumed physical isolation, disabling radio-layer encryption to reduce complexity and latency. However, as our proof-of-concept demonstrates, this design choice constitutes a critical vulnerability, as it leaves application data flows(e.g. OCPP transactions) completely exposed and vulnerable to eavesdropping via the F1-U midhaul interface.
\end{compactenum}

\begin{figure}[!htbp]
    \centering
    \includegraphics[width=\linewidth]{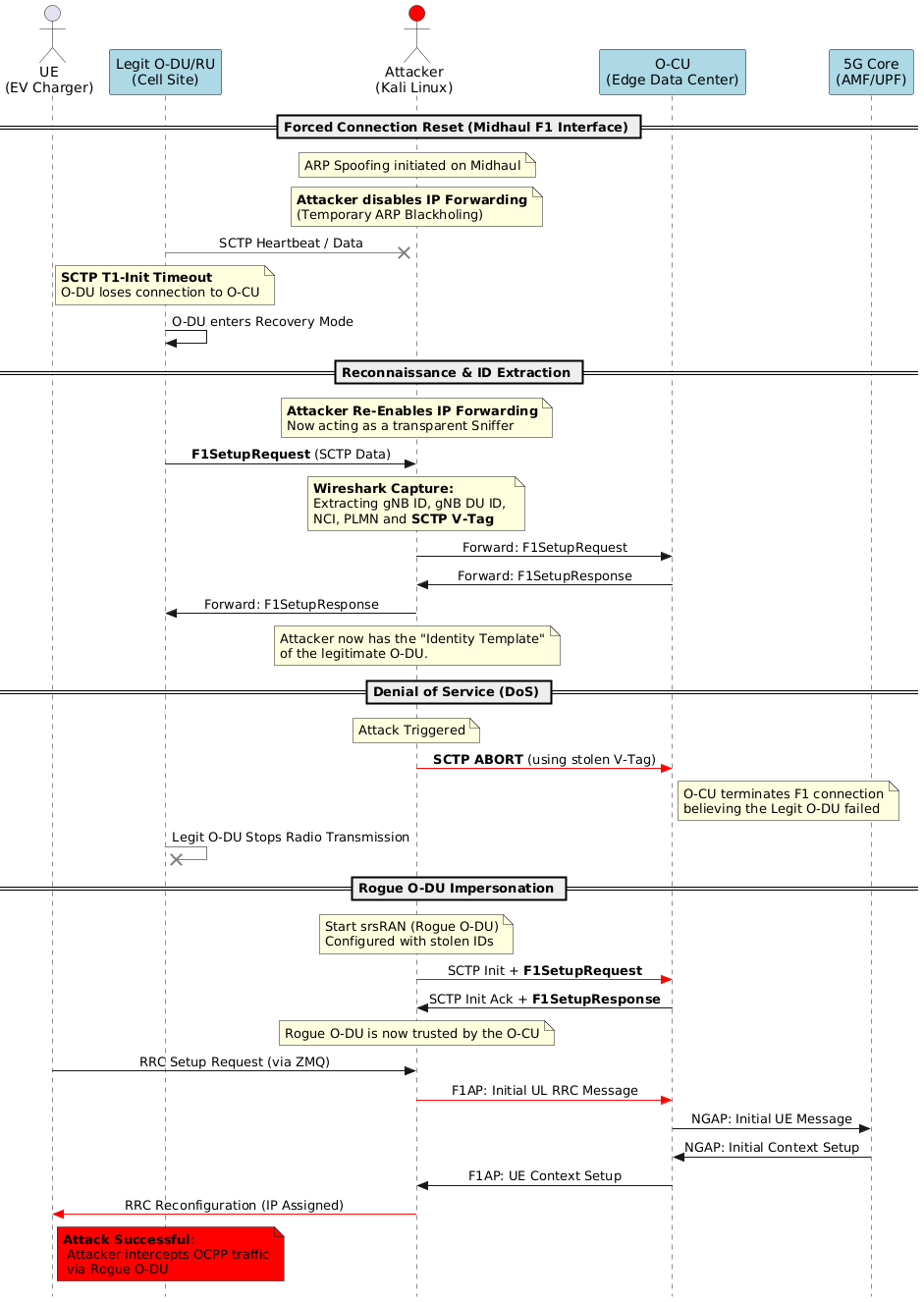} 
    \caption{The proposed threat model and attack methodology, including the targeted SCTP Abort and Rogue O-DU insertion on the F1 interface.}
    \label{fig:uml_attack}
\end{figure}

After the rogue O-DU is established, the attacker can observe the traffic forwarded through the compromised path. In the evaluated configuration enables extraction of plaintext OCPP messages, including fields such as \textit{idTag} and metering data. Although an attacker controlling this position could attempt message modification, such integrity manipulation was not experimentally evaluated in this work. The complete sequence of the proposed F1 midhaul spoofing attack, from the initial network reconnaissance to the successful rogue base station impersonation is depicted in Fig. \ref{fig:uml_attack}.
\vspace{-.5cm}
\vspace{-.1cm}

\subsection{Attacker Model and Scope}
The attacker is assumed to have obtained access to the IP midhaul segment connecting the O-CU and O-DU. This access may result from compromise of a virtual-machine host, a misconfigured virtual switch, insufficient network segmentation, insider access, or compromise of another node connected to the same transport network. The attacker can passively observe F1 traffic, perform ARP spoofing, inject forged SCTP packets, and launch a preconfigured rogue O-DU instance.

The attack does not assume compromise of the 5G Core, subscriber credentials, or the legitimate O-CU and O-DU software binaries. The evaluated scenario further assumes that F1-C and F1-U transport protection is not enabled and that the attacker can observe or inject packets on the midhaul. The plaintext OCPP-interception result additionally assumes that PDCP user-plane ciphering is disabled. Therefore, this paper evaluates the security consequences of an unprotected or misconfigured private O-RAN deployment. It does not claim that the proposed attack succeeds against deployments that enforce authenticated and encrypted F1 transport, strong CU--DU identity validation, and PDCP ciphering.

\subsection{Reconnaissance and Parameter Extraction}
F1 Application Protocol (F1AP) messages are encapsulated in SCTP packets. During the initialization phase, the attacker uses a packet sniffer (e.g., Wireshark), on the midhaul network to monitor traffic. The attacker successfully captures the \textit{F1SetupRequest} message in plain text format transmitted by the legitimate O-DU during startup. This step allows extracting the mandatory authentication parameters required by the O-CU, including 1) the globally unique \textit{gNB ID}, 2) the \textit{gNB DU ID}, 3) the Public Land Mobile Network \textit{PLMN} identity \textit{MCC} and \textit{MNC}, 4) the NR Cell Identity \textit{NCI} of the served cells.

\subsection{Targeted Denial of Service (SCTP Abort)}
To seamlessly replace the legitimate O-DU, the attacker must first terminate its active connection with the O-CU. This is achieved through a targeted DoS attack at the transport layer. The attacker injects a falsified \textit{SCTP ABORT} packet addressed to the O-CU, spoofing the source IP address to match that of the legitimate O-DU \cite{moheddine2025identifying, hussain2019privacy}. Assuming that the actual O-DU antenna has failed, the O-CU immediately terminates the SCTP connection and deletes the F1 frame for that DU.

\subsection{F1 Spoofing and Introduction of a Rogue DU}
Exploiting the short time window during which the O-CU is actively awaiting reconnection, the attacker creates an illegal O-DU. Using the parameters extracted during the reconnaissance phase, as \textit{gNB ID, PLMN, NCI}, the Rogue O-DU transmits a falsified \textit{F1SetupRequest}. Because the provided parameters are verified by the O-CU's configuration, the CU authenticates the connection and responds with a \textit{F1SetupResponse}. Consequently, the legitimate O-DU disconnects and the rogue O-DU establishes an F1 association in the evaluated environment. This  disrupts the legitimate service and creates a path for observing control- and user-plane traffic.

\section{6G Emulation Testbed Design}

The proposed testbed is designed to emulate a 6G-oriented radio access and core
network environment, grounded in the 3GPP Release~15 5G New Radio Standalone
(NR~SA) specification~\cite{3gpp_ts38300} and extended to reflect architectural
principles anticipated in 6G deployments, including disaggregated RAN, open
interfaces, and near-RT RAN intelligence~\cite{itu_imt2030}. All network
functions are instantiated as software processes running on isolated Virtual
Machines (VMs). VMs are hosted on a single physical host using Oracle VirtualBox with a
NAT Network configuration, assigning each node a static IPv4 address within the
\textit{192.168.100.0/24} subnet. Such fully virtualized approach enables
reproducible, hardware-independent experimentation while preserving
fidelity across all standard interfaces~\cite{bonati2020openran}.

\subsection{Design Principles}
The architecture of the simulation test environment used in this work is based on three main areas: virtualization, modularity, and reproducibility. By fully utilizing virtualized components and open-source implementations such as Open5GS and the srsRAN Project, the test environment eliminates the need for expensive, proprietary radio frequency (RF) hardware and spectrum licenses. Furthermore, the modular architecture strictly adheres to 3GPP and O-RAN decomposition standards \cite{3gpp_ts38470, oran_wg3_e2}, explicitly isolating the O-CU and O-DU into separate virtual environments. This design choice provides a highly observable, controlled and secure environment, which is essential for collecting plaintext traffic and analyzing targeted attacks at the F1 interface without causing real disruptions.
 
\subsection{Network Topology}
\begin{figure}[!htbp]
\centerline{\includegraphics[width=\linewidth]{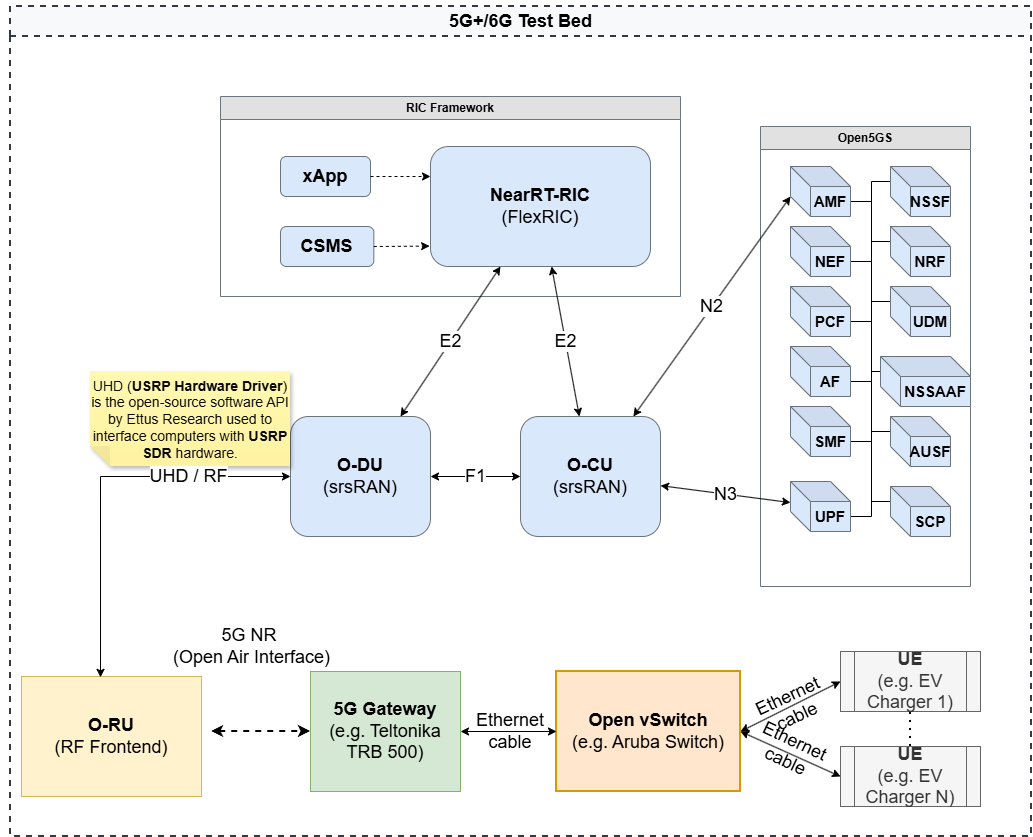}}
\caption{Topological diagram of the 6G-oriented emulated network testbed, highlighting
the spoofed gNB.}
\label{fig:topology}
\end{figure}

The testbed (Fig.~\ref{fig:topology}) comprises of five VMs, each fulfilling a distinct role in
the emulated network. The core
network functions are hosted on VM\,1 (\textit{192.168.100.4}), running the
Open5GS suite~\cite{open5gs}, which provides the Access and Mobility Management
Function (AMF) and the User Plane Function (UPF). The AMF communicates with the
RAN over the N2 interface using the NGAP protocol over SCTP, while the UPF
anchors the user-plane data path via the N3 GTP-U interface, both as defined in
3GPP TS~38.413~\cite{3gpp_ts38413} and TS~29.281~\cite{3gpp_ts29281},
respectively. 
The RAN is realized through a disaggregated gNB deployment following the 3GPP
F1 functional split defined in TS~38.470~\cite{3gpp_ts38470}. The Central Unit
(CU), hosted on VM\,2 (\textit{192.168.100.17}), is further subdivided into the
CU Control Plane (CU-CP) and the CU User Plane (CU-UP), which communicate
internally over the E1 interface~\cite{3gpp_ts37483}. The CU-CP connects to the
AMF over N2 and exposes an F1-C interface on port~38472 (SCTP) for
control-plane signaling toward the Distributed Unit. The CU-UP anchors the
GTP-U user-plane tunnel toward the UPF over N3 (UDP port~2152).
 
The Distributed Unit (DU), hosted on VM\,3 (\textit{192.168.100.18}), terminates
the F1 interface toward the CU and manages the lower layers of the RAN protocol
stack, including scheduling, HARQ, and the physical layer. Rather than a
physical radio front-end, the DU employs the ZeroMQ (ZMQ) software-defined RF
transport layer as a virtual air interface~\cite{srsran_project_github},
streaming downlink I/Q samples over TCP port~2000 and receiving uplink samples
on TCP port~2001.
 
The User Equipment is emulated on VM\,4 (\textit{192.168.100.19}) using
srsUE~\cite{gomez2016srslte}, the UE component of the srsRAN~4G suite
developed by Software Radio Systems\footnote{https://docs.srsran.com/projects/project/en/latest/}. Upon successful attach,
the UE is assigned an IP address from the \textit{10.45.0.0/16} pool managed by
the UPF and routed through a dedicated Linux network namespace (\textit{ue1}).
 
RAN intelligence is introduced through VM\,5 (\textit{192.168.100.20}), which
hosts the FlexRIC Near-RT RAN Intelligent Controller~\cite{schmidt2021flexric}.
The DU connects to the RIC over the E2 interface (SCTP port~36421), exposing
telemetry and control hooks via the E2SM-KPM and E2SM-RC service models,
enabling near-real-time monitoring of per-UE and cell-level key performance
metrics in accordance with the O-RAN Alliance E2 specification~\cite{oran_wg3_e2}.
 
The security threat is introduced through an additional VM running a Kali Linux
environment 
that acts as a software-defined rogue base station. The spoofed gNB is
configured to broadcast on the same frequency band as the legitimate DU, emitting
a counterfeit cell with the intent of luring the UE away from the authentic
network~\cite{yang2024fakebs,shaik2016practical}. The attack exploits the
absence of mutual authentication at the physical and RRC layers prior to the
establishment of a security context~\cite{rupprecht2018security,basin2018formal},
a known vulnerability that remains a subject of active research in the context
of 6G security design~\cite{itu_imt2030}. This node operates independently of
the legitimate RAN infrastructure and has no connection to the core network,
representing a realistic out-of-network adversarial presence. The placement of
the spoofed gNB within the topology is depicted in Fig.~\ref{fig:topology}.
 
Furthermore, each node in the testbed is configured with software and parameters specifically
chosen to maintain protocol correctness across all interfaces while enabling the
security scenario under investigation.

\subsubsection{5G Core Network through Open5GS (VM\,1)}
 
The core network is implemented using Open5GS~\cite{open5gs}, an open-source
3GPP-compliant 5G SA core. The AMF is bound to \textit{192.168.100.4} on the
NGAP interface and configured with PLMN \textit{99970} (MCC\,=\,999,
MNC\,=\,70), TAC\,=\,1, and two S-NSSAI entries corresponding to SST\,=\,1
with slice descriptors \textit{0x010203} and \textit{0x020304}, as defined in
3GPP TS~23.501~\cite{3gpp_ts23501}. The UPF is co-located on the same VM and
manages the \textit{10.45.0.0/16} UE address pool under the \textit{internet}
DNN. Subscriber provisioning is performed via the Open5GS WebUI, with the UE's
IMSI, authentication key~$K$, and operator key~$OPc$ pre-registered in the
Unified Data Repository (UDR) prior to the experiment.
 
\subsubsection{Central Unit through srsRAN CU (VM\,2)}
 
The CU is instantiated using the srsRAN Project open-source
implementation~\cite{srsran_project_github}, specifically commit
\textit{4bf1543936} on the \textit{release\_24\_10} branch. The CU-CP is
configured to connect to the AMF at \textit{192.168.100.4:38412} and to listen
for F1-C associations from the DU on \textit{192.168.100.17:38472}. The CU-UP
binds its F1-U GTP-U endpoint to the alias address
\textit{192.168.100.171:2152}, separating it from the N3/NGU interface bound to
\textit{192.168.100.17:2152} to avoid socket conflicts on the shared UDP
port~2152. Both the CU-CP and CU-UP connect to the FlexRIC Near-RT RIC over the
E2 interface~\cite{oran_wg3_e2,schmidt2021flexric}, enabling near-real-time
RAN control and KPM collection from the centralised unit.
 
\subsubsection{Distributed Unit through srsRAN DU (VM\,3)}
 
DU and CU are built using the same srsRAN version~\cite{srsran_project_github}
and deployed on a separate VM realizing the physical CU/DU split over an IP
network~\cite{3gpp_ts38470}. The cell is configured on NR band~3 (FDD) with
downlink ARFCN 368500, corresponding to 1842.5\,MHz center frequency, 10\,MHz channel bandwidth, and 15\,kHz subcarrier spacing. This
configuration is a deliberate design choice: the srsUE implementation of NR SA
mode is constrained to 15\,kHz subcarrier spacing, limiting operation to FDD
bands~\cite{gomez2016srslte}. The physical cell identity is set to PCI\,=\,1
and SSB ARFCN is 367930.
 
The RF front-end is replaced by a ZMQ transport layer~\cite{srsran_project_github}
operating at a base sample rate of 11.52\,MHz. The DU binds its ZMQ sockets on
\textit{0.0.0.0} (all interfaces) to ensure cross-VM connectivity within the
VirtualBox NAT network. TX and RX gains are set to 89\,dB to ensure sufficient
signal-to-noise ratio for reliable Msg3 decoding during the random access
procedure~\cite{3gpp_ts38321}. The DU also establishes an E2 connection to the
FlexRIC RIC at \textit{192.168.100.20:36421}, advertising E2SM-KPM and E2SM-RC
service model support~\cite{schmidt2021flexric,oran_wg3_e2}.
 
\subsubsection{User Equipment through srsUE (VM\,4)}
 
The UE is emulated using srsRAN~4G~\cite{gomez2016srslte}, supporting NR SA
operation as a dual-mode client. The UE is configured to search for an NR cell
on band~3, with \textit{dl\_nr\_arfcn\,=\,368500},
\textit{ssb\_nr\_arfcn\,=\,367930}, and both data and SSB subcarrier spacing set
to 15\,kHz. The USIM is provisioned with IMSI \textit{999700000000001},
authentication algorithm Milenage~\cite{3gpp_ts35206}, and the corresponding
$K$ and $OPc$ values matching the Open5GS subscriber record. Network slicing is
enabled with NSSAI SST\,=\,1, SD\,=\,66051 (decimal representation of
\textit{0x010203}), consistent with the AMF and CU slice
configuration~\cite{3gpp_ts23501}. Upon attach, the UE traffic is isolated in a Linux network namespace (\textit{ue1}) via a \textit{tun\_srsue}
tunnel interface.
 
\subsubsection{Near-RT RIC through FlexRIC (VM\,5)}
 
The Near-RT RAN Intelligent Controller is deployed using the FlexRIC
framework~\cite{schmidt2021flexric}, which implements the O-RAN E2 interface
specification~\cite{oran_wg3_e2} and supports the E2SM-KPM and E2SM-RC service
models. FlexRIC listens for E2 node connections on SCTP port~36421 and, upon
successful E2 Setup with both the CU and DU, enables near-real-time collection
of cell and UE-level KPMs including throughput, BLER, and scheduling metrics. In this experiment, RIC serves as a passive monitoring plane,
providing visibility into normal network behavior as a baseline against which
anomalous conditions introduced by the rogue gNB can be contextualized.
 
\subsubsection{Attacker Node and Rogue O-DU Emulation}



To evaluate the security of the disaggregated architecture and execute the proposed F1 spoofing attack, a dedicated adversary node was integrated into the emulation testbed. This node is hosted on a separate virtual machine (VM6) running the Kali Linux operating system. Strategically positioned within the midhaul IP network, it shares the same subnet as the legitimate O-CU and O-DU, allowing it to act as a Man-in-the-Middle (MitM) \cite{aiello2026investigating}. The attacker system is equipped with a suite of network analysis and exploitation tools. Specifically, Wireshark is utilized for passive packet sniffing to capture unencrypted F1 Application Protocol (F1AP), and SCTP traffic. Furthermore, packet manipulation scripts are deployed to forge and inject SCTP network packets \cite{moheddine2025identifying}. To achieve the final stage of the attack, an instance of the srsRAN Project software is installed on the attacker system. This instance is pre-configured to operate as a malicious Rogue O-DU \cite{saedi2020generation, xu2025integrity}, enabling the attacker to leverage the intercepted parameters to actively hijack the base station's F1 interface, while utilizing ZeroMQ (ZMQ) to establish a virtualized radio link with the emulated  EV Charger (UE), thereby completing the end-to-end interception path. This approach exploits the lack of strict mutual authentication during the initial setup phases, a vulnerability documented in state-of-the-art \cite{shaik2016practical}.

\section{Experimental Results and Evaluation}
This section aims to evaluate the practical execution of the proposed F1 interface spoofing attack within the virtualized O-RAN testbed. By sequentially deploying the attack phases described in Section II we demonstrate service disruption and, plaintext exposure of EV-charging application traffic. These outcomes apply to the evaluated unprotected F1 configuration and should not be interpreted as applying to deployments that enforce the relevant transport and RAN security protections.

\subsection{Malicious Control Plane Establishment}
\begin{figure}[!htbp]
    \centering
    \includegraphics[width=\linewidth]{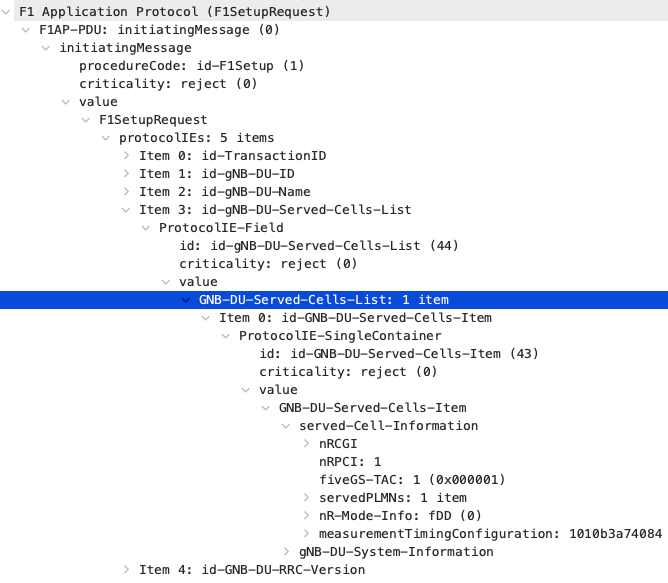} 
    \caption{Packet capture illustrating critical F1AP parameters.}
    \label{fig:wireshark}
\end{figure}

The attack is initiated from the attacker system Kali Linux (VM6) by launching the ARP spoofing and SCTP Abort scripts against the legitimate O-CU/O-DU F1 link. Consequently, the legitimate O-DU loses its midhaul connection and ceases radio transmissions. Exploiting this brief outage window, the pre-configured Rogue O-DU instance is launched using the srsRAN Project software. As depicted in Fig. \ref{fig:wireshark}, the attacker utilizes Wireshark to passively monitor the midhaul link and extract critical parameters from the unencrypted F1SetupRequest message, as \textit{id-TransactionID}, \textit{id-gNB-DU-ID}, \textit{id-gNB-DU-Name}, \textit{id-gNB-DU-Served-Cells-List}, \textit{id-GNB-DU-RRC-Version}.

\begin{figure}[!htbp]
    \centering
    \includegraphics[width=\linewidth]{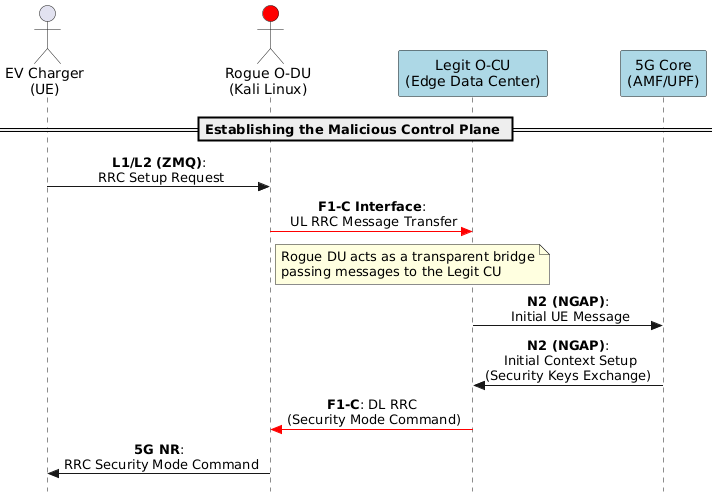} 
    \caption{The Rogue O-DU (Kali Linux) successfully bridges the RRC connection via ZMQ and forwards the signaling to the legitimate O-CU.}
    \label{fig:MaliciousControlPlane}
\end{figure}

Since the Rogue O-DU utilizes the intercepted identifiers \textit{gNB ID, NCI}, the legitimate O-CU accepts the incoming \textit{F1SetupRequest} without triggering any security alarms. Subsequently, the virtual EV Charger (UE) detects the rogue gNB via the ZeroMQ (ZMQ) radio link and initiates an RRC connection, mistaking it for the legitimate infrastructure. Initially, the Rogue O-DU acts as a transparent bridge, forwarding the \textit{NAS/NGAP} messages to the Core while maintaining full control over the lower layers of the connection. Fig. \ref{fig:MaliciousControlPlane} illustrates the exact signaling sequence of this phase. 

\subsection{User Plane Interception and Data Compromise}
\begin{figure}[!htbp]
    \centering
    \includegraphics[width=\linewidth]{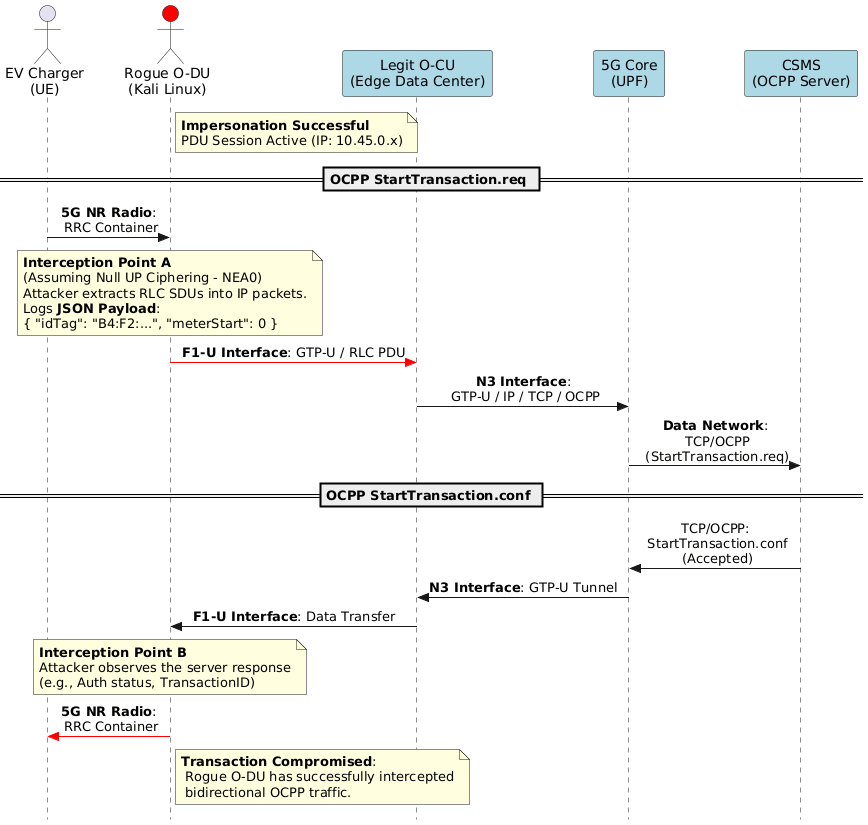} 
    \caption{Operating under NEA0, the attacker successfully extracts plaintext OCPP payloads at the F1-U interface.}
    \label{fig:TransactionviaRoguegNB}
\end{figure}

Once the malicious control plane is established, the 5G Core assigns an active PDU session \textit{IP: 10.45.0.x} to the EV charger, enabling communication with the CSMS. This phase represents the most critical part of the experiment, namely the interception of the OCPP application data. The testbed uses the NEAO \cite{3gpp_ts33501}, to accurately emulate an unencrypted private 5G network environment. As the EV charger transmits  \textit{StartTransaction.req} message over the 5G network, the Rogue O-DU captures the F1-U traffic. Fig. \ref{fig:TransactionviaRoguegNB} depicts the interception sequence at the user plane. Due to the effective disabling of the PDCP layer encryption, the Radio Link Control (RLC) Service Data Units (SDUs) are extracted, and the underlying IP packets are reassembled in plaintext through Wireshark.

\subsection{Impact Analysis}
In this section, we demonstrate confidentiality exposure and availability disruption, hence integrity attacks are a plausible consequence but were not experimentally validated. By capturing the plaintext JSON payloads, sensitive EV charger parameters, such as the \textit{idTag}, \textit{meterValues}, \textit{meterStop} values, are successfully extracted. In a real-world deployment, an attacker controlling a compromised traffic path may attempt to alter metering values or inject OCPP commands. However, these integrity and command-injection scenarios were not tested in the present experiment and are therefore identified as security implications and future evaluation targets rather than demonstrated outcomes \cite{11130009}.

\section{Conclusion}
This paper presented a security evaluation of a 6G-oriented O-RAN architecture through the design and execution of a targeted F1 midhaul spoofing attack. By exploiting the trust between the O-CU and O-DU, the proposed methodology demonstrated how an adversary with access to the midhaul network can extract critical F1AP parameters, disrupt the legitimate O-DU connection through ARP spoofing and SCTP Abort, and introduce a rogue O-DU into the network. The findings highlight that mutual authentication between RAN components, security in F1-C and F1-U traffic, and continuous monitoring of midhaul behavior are essential for reducing rogue component insertion and traffic interception.

Future work focuses on additional attack scenarios, including rogue O-RU insertion, E2 interface manipulation, and adversarial interaction with Near-RT RIC applications. Moreover, the testbed will be enhanced with stronger detection and mitigation mechanisms as the integration of anomaly detection techniques for SCTP and F1AP traffic, monitoring of unexpected O-DU reconnection events, and validation of RAN component identities before accepting F1 associations. The role of the Near-RT RIC in detecting abnormal control-plane and user-plane behavior will also be examined.

\printbibliography 

\end{document}